\documentclass{article}
\usepackage{arxiv}
\usepackage[utf8]{inputenc}
\usepackage[T1]{fontenc}
\usepackage{lmodern}
\usepackage{amsmath,amssymb,amsthm,mathtools}
\usepackage{array,booktabs,tabularx}
\usepackage{xcolor,graphicx}
\usepackage{tikz}
\usetikzlibrary{arrows.meta,positioning,calc,fit,shapes.geometric}
\usepackage{algorithm}
\usepackage{algpseudocode}
\usepackage{url}
\usepackage[font=small,labelfont=bf]{caption}
\usepackage{subcaption}
\usepackage{microtype}
\usepackage[section]{placeins}
\usepackage[hidelinks]{hyperref}

\definecolor{navylight}{HTML}{E8E8E8}
\definecolor{navymid}{HTML}{747474}
\definecolor{navydark}{HTML}{303030}
\definecolor{forestlight}{HTML}{DCDCDC}
\definecolor{forestmid}{HTML}{666666}
\definecolor{forestdark}{HTML}{282828}
\definecolor{burgundylight}{HTML}{F1F1F1}
\definecolor{burgundymid}{HTML}{8A8A8A}
\definecolor{burgundydark}{HTML}{424242}
\definecolor{camellight}{HTML}{E4E4E4}
\definecolor{camelmid}{HTML}{7C7C7C}
\definecolor{cameldark}{HTML}{383838}
\definecolor{graylight}{HTML}{FFFFFF}
\definecolor{graymid}{HTML}{8C8C8C}
\definecolor{graydark}{HTML}{2F2F2F}

\colorlet{purplelight}{burgundylight}
\colorlet{purplemid}{burgundymid}
\colorlet{purpledark}{burgundydark}
\colorlet{teallight}{forestlight}
\colorlet{tealmid}{forestmid}
\colorlet{tealdark}{forestdark}
\colorlet{corallight}{burgundylight}
\colorlet{coralmid}{burgundymid}
\colorlet{coraldark}{burgundydark}
\colorlet{amberlight}{camellight}
\colorlet{ambermid}{camelmid}
\colorlet{amberdark}{cameldark}
\colorlet{bluelight}{navylight}
\colorlet{bluemid}{navymid}
\colorlet{bluedark}{navydark}

\providecommand{\dtype}{d_{\mathrm{type}}}
\providecommand{\dtax}{d_{\mathrm{tax}}}
\providecommand{\dangle}{d_{\angle}}
\providecommand{\dfield}{d_{\mathrm{field}}}
\providecommand{\dnode}{d_N}
\providecommand{\dw}{d_{\mathrm{wf}}}
\providecommand{\dwh}{\widehat d_{\mathrm{wf}}}

\tikzset{
  wfnode/.style={draw, rounded corners=3pt, minimum width=2.65cm,
    minimum height=0.62cm, font=\small\bfseries, inner sep=3pt, align=center},
  nd-trig/.style={wfnode, fill=forestlight, draw=forestmid, text=forestdark},
  nd-out/.style={wfnode, fill=navylight, draw=navymid, text=navydark},
  nd-inp/.style={wfnode, fill=camellight, draw=camelmid, text=cameldark},
  nd-neut/.style={wfnode, fill=graylight, draw=graymid, text=graydark},
  nd-ins/.style={wfnode, fill=burgundylight, draw=burgundymid,
    text=burgundydark, dashed},
  nd-lc/.style={wfnode, fill=burgundylight, draw=burgundymid,
    text=burgundydark},
  trnd/.style={draw, rounded corners=3pt, minimum width=2.0cm,
    minimum height=0.50cm, font=\scriptsize\bfseries, inner sep=2pt, align=center},
  trnd-root/.style={trnd, fill=graylight, draw=graymid, text=graydark},
  trnd-trig/.style={trnd, fill=forestlight, draw=forestmid, text=forestdark},
  trnd-out/.style={trnd, fill=navylight, draw=navymid, text=navydark},
  trnd-inp/.style={trnd, fill=camellight, draw=camelmid, text=cameldark},
  trnd-lc/.style={trnd, fill=burgundylight, draw=burgundymid,
    text=burgundydark, minimum width=2.8cm},
  trnd-tp/.style={trnd, minimum width=1.85cm, font=\scriptsize\bfseries},
  trnd-tp-trig/.style={trnd-tp, fill=forestlight, draw=forestmid, text=forestdark},
  trnd-tp-out/.style={trnd-tp, fill=navylight, draw=navymid, text=navydark},
  trnd-tp-lc/.style={trnd-tp, fill=burgundylight, draw=burgundymid,
    text=burgundydark},
  trnd-tp-inp/.style={trnd-tp, fill=camellight, draw=camelmid, text=cameldark},
  tree-e/.style={draw=graymid!75, line width=0.60pt},
  arr-id/.style={-{Stealth[length=4pt]}, draw=forestmid, line width=1.2pt},
  arr-sub/.style={-{Stealth[length=4pt]}, draw=cameldark, line width=1.1pt, dashed},
  wfedge/.style={-{Stealth[length=3pt]}, draw=graydark!75, line width=0.65pt},
  wfedge-disc/.style={-{Stealth[length=3pt]}, draw=burgundydark,
    line width=0.85pt, dashed}
}

\hypersetup{
  pdftitle={A Metric for Semantic Graph Matching of No-Code Workflows},
  pdfauthor={Bartlomiej Matejczyk and Joao Tiago Aparicio}
}
\newcolumntype{Y}{>{\raggedright\arraybackslash}X}

\theoremstyle{plain}
\newtheorem{theorem}{Theorem}[section]
\newtheorem{lemma}[theorem]{Lemma}

\theoremstyle{definition}
\newtheorem{definition}[theorem]{Definition}
\newtheorem{assumption}[theorem]{Assumption}
\theoremstyle{remark}

\providecommand{\M}{\mathcal{M}}
\providecommand{\T}{\mathcal{T}}
\providecommand{\K}{\mathcal{K}}
\providecommand{\R}{\mathbb{R}}
\providecommand{\dnodebar}{\bar d_N}
\providecommand{\missing}{\bot}
\providecommand{\dom}{\operatorname{dom}}
\providecommand{\im}{\operatorname{im}}
\providecommand{\Description}[1]{}

\title{A Metric for Semantic Graph Matching \\ 
of No-Code Workflows}

\author{
  Bartłomiej Matejczyk
  \hspace{2.5em}
  João Tiago Aparício\textsuperscript{1}
  \\[0.6em]
  {\normalsize\textsuperscript{1}INESC-ID, IST, Universidade de Lisboa}
  \\[-0.05em]
  {\normalsize Lisboa, Portugal}
  \\[0.15em]
  {\small\texttt{joao.aparicio@tecnico.ulisboa.pt}}
}

\date{}

\begin{document}
\maketitle

\begin{abstract}
No-code and agentic systems increasingly produce executable workflows, but a
workflow is not adequately compared as a string: its typed operations,
parameters, and directed connections jointly determine the artifact.  We define
an exact semantic graph-matching distance for finite attributed workflow graphs.
One partial one-to-one node matching determines both node substitution and
directed labelled edge disagreement, while unmatched nodes pay a positive,
configuration-sensitive null cost.  Under a frozen type hierarchy, encoder,
field universe, and metric-compatible primitive distances, the resulting
objective is a metric on workflow graphs up to attributed isomorphism.  A padded
Hungarian procedure provides a scalable feasible upper bound while remaining
explicitly separate from the exact objective.

We implement and evaluate the proposed metric on 4,705 public n8n exports,
representing 2,282 distinct persisted normalized-graph hashes, 73,057 nodes, and
469 observed node types.  Across 32,093 branch-and-bound-complete Experiment~1 evaluations, the
normalized-graph equal-weighted mean raw distance is near zero under
surface-preserving transformations ($1.21\times10^{-5}$) and increases to
$0.584$ under major type or subgraph changes.  On a complete 796-hash panel,
accumulated edit count has Spearman $\rho=0.520$ with distance, while the median
within-workflow correlation is $0.975$.  One- and three-perturbation systems are
clearly separated, with marginal means of $0.307$ and $0.763$, respectively, and
all 117,312 evaluated triangle orientations have non-negative slack.  Weight
ablations preserve the positive edit-count association across all tested
configurations.  Together, these controlled tests provide convergent evidence of numerical
consistency, near-invariance to surface-preserving transformations, graded
sensitivity to semantic and structural change, and robust discrimination across
controlled workflow modifications.  The metric is
an artifact-level measure and does not by itself assert behavioural equivalence
under external state or runtime data.
\end{abstract}

\section{Introduction}

Low-code and no-code platforms turn integration logic into executable visual
artifacts.  A typical workflow combines triggers, service calls, transformations,
conditions, and language-model tools through a directed connection map.  Recent
systems reuse or generate such structures from examples and natural-language
requirements \cite{du2026swift,zhang2026reusstdflow}.  At the same time, empirical
work has begun to characterize thousands of public n8n agentic workflows and
their reliability patterns \cite{tang2026n8n}.  Generation, retrieval, and corpus
analysis all create the same measurement problem: when are two workflow
artifacts close, and which disagreement makes them different?

The serialized JSON is a poor measurement space.  Cosmetic names and layout
coordinates may change while execution-relevant structure remains fixed.  The
opposite failure is equally important: two exports can mention the same services
and parameters while routing data through different branches.  A useful
artifact-level comparison must establish node correspondence, charge semantic
and parameter differences under that correspondence, and then measure the
remaining topology disagreement.

This paper develops such a distance.  A workflow is represented as a finite
attributed directed graph whose nodes carry registry types and canonical
parameter dictionaries and whose edges preserve direction and connection labels.
The distance minimizes a single objective over partial one-to-one node
matchings.  Matched nodes pay type-and-field substitution costs, unmatched nodes
pay strictly positive null-edit costs, and topology is compared after both
workflows are lifted through the same correspondence.  This yields a transparent
workflow-specific graph edit metric in which semantic and structural
disagreement are evaluated jointly rather than through independent scores.

The work makes four contributions.

\begin{enumerate}
  \item It specifies a canonical graph representation for no-code workflows,
  including type, parameter-field, null-edit, and directed labelled-edge
  semantics under a frozen comparison context.
  \item It gives explicit sufficient conditions under which the joint
  edge-aware objective is a metric, with a complete proof in
  Appendix~\ref{app:proof}.
  \item It provides an exact branch-and-bound procedure and a padded Hungarian
  approximation whose relationship to the exact objective is formally stated.
  \item It validates the metric on a large public workflow corpus through
  representation-preserving transformations, graded semantic and structural
  perturbations, accumulated edits, weight ablations, system-level
  discrimination, and an empirical triangle audit.
\end{enumerate}

A dependable distance supports more than leaderboard scoring.  It makes
workflow-generator regression tests sensitive to both semantic and routing
errors, identifies near-duplicates before a template library grows, retrieves a
structurally compatible starting workflow from a corpus, and audits whether a
platform migration preserved an automation's executable shape.  Because each
score decomposes into matched-node, null-edit, and edge terms, the measurement is
also diagnostically useful rather than merely rank-producing.

The scope is artifact-level agreement.  The metric characterizes similarity of
the represented workflow structure and attributes; it does not replace runtime
verification when credentials, external state, service versions, or input data
are part of the question.  Trajectory-level policy evaluation, such as Near-Miss
\cite{rabinovich2026nearmiss}, addresses that complementary behavioural level.

\section{Related work and contribution boundary}

\paragraph{Graph edit distance.}
The direct mathematical lineage is graph edit distance (GED), which compares
attributed graphs through minimum-cost edit paths
\cite{sanfeliu1983distance,serratosa2021redefining}.  Exact formulations include
binary optimization and exact search \cite{justice2006binary,blumenthal2017exact};
larger instances commonly use assignment-based or other heuristics
\cite{zeng2009comparing,blumenthal2020heuristics}.  In particular, node-map and
quadratic-assignment formulations already use one correspondence to induce both
node and edge edits \cite{bougleux2017qap}.  We therefore do not claim the shared
correspondence itself as a new graph-matching principle.  The contribution is a
workflow-specific metric construction: a frozen typed representation, explicit
parameter and null semantics, directed labelled-edge disagreement, and a complete
proof for their joint objective.  This also motivates a strict separation between
the exact objective and the Hungarian-first feasible upper bound used for scale.

\paragraph{Process and workflow similarity.}
Business-process retrieval already combines activity labels and topology.
Dijkman et al. formulate explicit graph matching for process-model search and
compare label, structural, and behavioural similarity
\cite{dijkman2009graph,dijkman2011similarity}.  Scientific-workflow retrieval has
also paired graph structure with domain attributes and cost constraints.  Ma et
al. prove a distance for cost-constrained scientific workflows
\cite{ma2015cost}, while Starlinger et al. study workflow similarity as a
retrieval problem across several representations and comparison functions
\cite{starlinger2014workflow}.  These are substantive precedents, not baselines
that this paper supersedes.  The present object is narrower in some respects (it does not
model behaviour) and more specific in others: node types, recursively extracted
parameter values, labelled directed connections, and null edits are evaluated
under one shared partial matching.

\paragraph{Semantic and learned graph comparison.}
Sentence encoders provide useful representations for short type descriptions
\cite{reimers2019sentencebert}, while learned graph-similarity models can provide
fast task-specific predictions \cite{bai2019simgnn}.  Neither alone supplies the
identity and triangle properties needed here.  We therefore use angular distance
between frozen type embeddings only as one term, protected by a positive
discrete identity term.  A learned model could later approximate the exact score,
but would require a separate error study.

\paragraph{No-code and agentic workflows.}
ReusStdFlow extracts reusable n8n workflow segments and combines graph and vector
retrieval; SWIFT transfers structural priors for agentic workflow generation
\cite{zhang2026reusstdflow,du2026swift}.  Tang et al. characterize the design and
reliability patterns of more than 6,000 public n8n agentic workflows
\cite{tang2026n8n}.  Those studies establish the relevance of reusable structure
and public workflow corpora, but they do not define the pairwise metric proposed
here.  Conversely, our perturbation study does not evaluate their generators.

The representation-first stance is consistent with work that makes network
encoding choices explicit and auditable \cite{aparicio2026networked}.  Related
work by Aparicio et al. also illustrates how a domain-specific distance can
combine network constraints with attributed temporal profiles
\cite{aparicio2023trips}; \texttt{tex2net} demonstrates a graph-native encoding
pipeline for unstructured artifacts \cite{aparicio2023tex2net}.  We cite these as
methodological precedents for explicit representation, not as prior workflow
distances.  Table~\ref{tab:prior_boundary} makes this contribution boundary
explicit.

\begin{table}[t]
\centering
\caption{Closest contribution boundaries.}
\label{tab:prior_boundary}
\begin{tabularx}{\textwidth}{@{} l Y Y @{} }
\toprule
Work & Main contribution & Boundary relative to this paper \\
\midrule
Dijkman et al. \cite{dijkman2009graph,dijkman2011similarity}
& Label, structural, and behavioural process-model similarity
& Process models and retrieval; no frozen typed-parameter metric of n8n exports. \\
Ma et al. \cite{ma2015cost}
& Metric retrieval for cost-constrained scientific workflows
& QoS- and datatype-oriented graph model rather than no-code registry fields. \\
ReusStdFlow \cite{zhang2026reusstdflow}
& Segment extraction, storage, and reuse for n8n workflows
& Generation and reuse framework, not a pairwise artifact metric. \\
Tang et al. \cite{tang2026n8n}
& Large-scale characterization of n8n agentic workflows
& Corpus-level patterns and reliability, not exact graph matching. \\
Near-Miss \cite{rabinovich2026nearmiss}
& Latent policy-failure detection from agent trajectories
& Behavioural policy adherence, complementary to static artifact distance. \\
\bottomrule
\end{tabularx}
\end{table}

\section{Canonical workflow graphs}\label{sec:representation}

\begin{definition}[Workflow graph]\label{def:workflow_graph}
A workflow graph is a tuple
\begin{equation}
G=(V,E,\tau,\alpha),
\end{equation}
where $V$ is a finite node set,
$E\subseteq V\times V\times\Lambda$ is a finite set of directed labelled
edges, $\tau:V\to\T$ maps nodes to fixed type identifiers, and
$\alpha:V\to\mathcal A$ maps nodes to canonical finite parameter dictionaries.
For simplicity, $(u,v,\ell),(u,v,\ell')\in E$ implies $\ell=\ell'$.
\end{definition}

The formal object is a directed labelled \emph{simple} edge set.  Repeated
connection records with identical endpoints and the same canonical label denote
the same structural relation and are represented once.  Connection type and
output-group information are incorporated in the label, so distinct routing
relations remain distinct.  An unlabelled platform can use a singleton label
set.  For n8n, the parser reads node objects, \texttt{parameters}, and the named
output groups in the connection map \cite{n8nexportdocs,n8ncomponentsdocs}.

The canonicalizer recursively sorts and flattens parameter dictionaries, keeps
list order, and removes display name, canvas position, and other top-level
presentation fields.  Nodes receive deterministic type-based identifiers in
topological order, with source order used only to stabilize cyclic
serializations.  These identifiers are serialization aids: the metric itself
optimizes over node matchings and is therefore independent of the original node
IDs.  Connection labels preserve the output name or connection type and its
index.  Parameter values are retained because they are part of the represented
workflow semantics; released aggregate results do not expose those values.

The persisted normalized-graph hash is a deterministic corpus-level grouping
key derived from the normalized serialization.  It is not an input to
$\dw$, and Theorem~\ref{thm:metric} does not depend on the hash being a
complete canonical-labelling procedure for graph isomorphism.

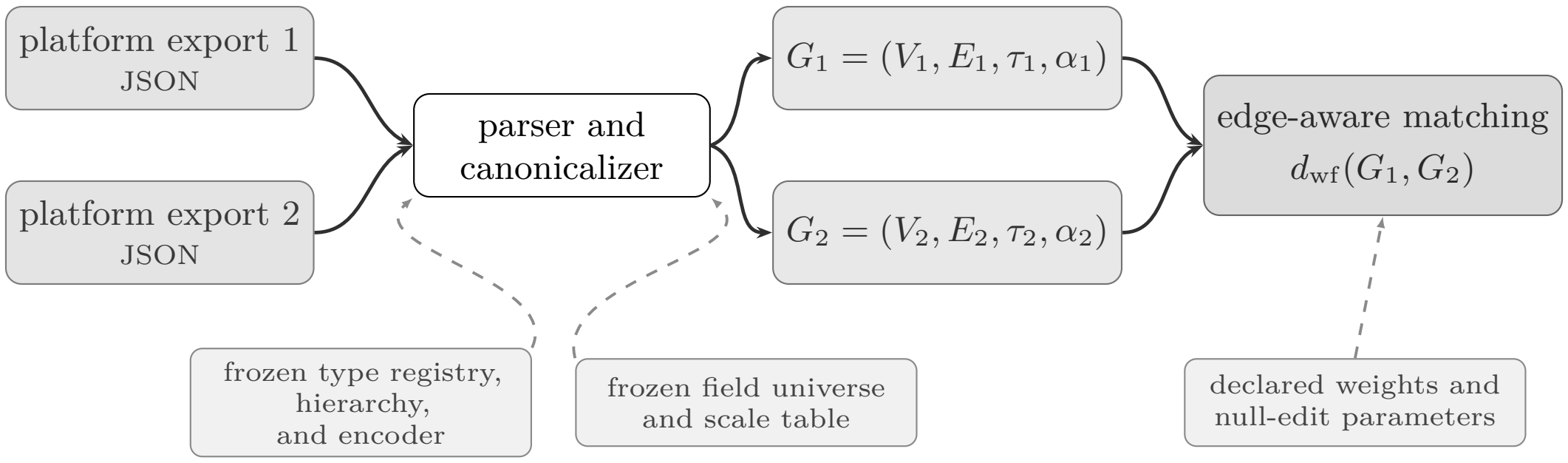
\begin{figure}[t]
\centering
\begin{tikzpicture}[
  x=1cm,y=1cm,
  stage/.style={draw, rounded corners=4pt, fill=graylight,
    minimum width=2.65cm, minimum height=0.92cm, align=center, font=\small},
  input/.style={stage, fill=camellight, draw=camelmid, text=cameldark},
  graph/.style={stage, fill=navylight, draw=navymid, text=navydark},
  metric/.style={stage, fill=forestlight, draw=forestmid, text=forestdark,
    minimum width=3.05cm, minimum height=1.25cm},
  resource/.style={draw, rounded corners=3pt, fill=burgundylight,
    draw=burgundymid, text=burgundydark, align=center,
    text width=2.75cm, minimum width=3.05cm, minimum height=0.78cm,
    font=\scriptsize},
  flow/.style={-{Stealth[length=4pt]}, draw=graydark, line width=0.90pt},
  feed/.style={-{Latex[length=3pt]}, draw=burgundymid, dashed, line width=0.72pt}
]
  \node[input] (j1) at (0,0.78) {platform export 1\\[-1pt]\scriptsize JSON};
  \node[input] (j2) at (0,-0.78) {platform export 2\\[-1pt]\scriptsize JSON};
  \node[stage] (canon) at (3.60,0) {parser and\\canonicalizer};
  \node[graph] (g1) at (7.05,0.78) {$G_1=(V_1,E_1,\tau_1,\alpha_1)$};
  \node[graph] (g2) at (7.05,-0.78) {$G_2=(V_2,E_2,\tau_2,\alpha_2)$};
  \node[metric] (dist) at (10.95,0) {edge-aware matching\\[2pt]
    $d_{\mathrm{wf}}(G_1,G_2)$};

  \draw[flow] (j1.east) to[out=0,in=155] (canon.west);
  \draw[flow] (j2.east) to[out=0,in=-155] (canon.west);
  \draw[flow] (canon.east) to[out=20,in=180] (g1.west);
  \draw[flow] (canon.east) to[out=-20,in=180] (g2.west);
  \draw[flow] (g1.east) to[out=0,in=155] (dist.west);
  \draw[flow] (g2.east) to[out=0,in=-155] (dist.west);

  \node[resource] (registry) at (1.80,-2.30)
    {frozen type registry,\\hierarchy, and encoder};
  \node[resource] (fields) at (5.25,-2.30)
    {frozen field universe\\and scale table};
  \node[resource] (weights) at (10.70,-2.30)
    {declared weights and\\null-edit parameters};
  \draw[feed] (registry.north east) to[out=65,in=-138] (canon.south west);
  \draw[feed] (fields.north west) to[out=105,in=-42] (canon.south east);
  \draw[feed] (weights.north) -- (dist.south);
\end{tikzpicture}
\caption{From two platform exports to one auditable comparison.  Both exports pass
through the same canonicalizer.  The registry, field scales, encoder, and metric
parameters are frozen parts of the measurement definition; changing any of them
defines a different distance.  Solid arrows preserve the two workflow inputs;
dashed arrows identify the frozen context used by the comparison.}
\Description{Two workflow exports pass through a common parser and canonicalizer to form attributed graphs, which are compared under a frozen registry, field universe, encoder, weights, and null-edit parameters.}
\label{fig:pipeline}
\end{figure}

The comparison context
\begin{equation}
\Theta=(\T,\mathcal H,e,\Lambda,c_\Lambda,\K_\star,
        \{c_k,\delta_k,\omega_k\}_{k\in\K_\star},\lambda,w,\beta,\gamma)
\end{equation}
is part of the measurement definition.  Here $\mathcal H$ is the type hierarchy,
$e$ the type-embedding table, $\Lambda$ and $c_\Lambda$ are the edge-label domain
and canonicalizer, $\K_\star$ is the nonempty field universe, and $c_k$, $\delta_k$,
and $\omega_k$ are typed field canonicalizers, metrics, and weights.  The
remaining symbols are type, node, edge, and null-edit weights.  The metric theorem
assumes $\Theta$ is frozen before any pair is scored.  Figure~\ref{fig:pipeline}
shows where these shared context objects enter the comparison pipeline.

\section{Ground distance on nodes}\label{sec:ground}

\subsection{Type distance}

Let $\mathcal H$ be a fixed connected hierarchy over the finite type set
$\T$.  With fixed diameter $D_{\mathcal H}>0$, define
\begin{equation}
\dtax(t,s)=\frac{d_{\mathcal H}(t,s)}{D_{\mathcal H}}.
\end{equation}
Let $e(t)\in\R^d\setminus\{0\}$ be a frozen description embedding and
\begin{equation}
\dangle(t,s)=\frac{1}{\pi}\arccos\!\left(
 \left\langle\frac{e(t)}{\lVert e(t)\rVert},
                    \frac{e(s)}{\lVert e(s)\rVert}\right\rangle\right).
\end{equation}
Angular distance is a metric on unit vectors and a pseudometric on type
identifiers when two identifiers share the same normalized embedding.  The
discrete term prevents that collapse.

\begin{definition}[Type distance]\label{def:type_distance}
For $\lambda_0,\lambda_1,\lambda_2\ge0$ such that
$\lambda_0>0$ and $\lambda_0+\lambda_1+\lambda_2=1$, let
\begin{equation}\label{eq:type_distance}
\dtype(t,s)=\lambda_0\mathbf{1}_{\{t\ne s\}}
 +\lambda_1\dtax(t,s)+\lambda_2\dangle(t,s).
\end{equation}
\end{definition}

The positive discrete metric supplies identity of indiscernibles; the other
terms are pseudometrics, so their nonnegative weighted sum is a metric and lies
in $[0,1]$.  The implementation uses
$\lambda_0=\lambda_1=\lambda_2=1/3$.  Types absent from the registry are
attached to a common root before evaluation, preserving a total and
pair-independent type domain.  The numerical implementation uses frozen
finite-precision embeddings; the resulting self-comparison residual is reported
explicitly in the invariance experiment and is negligible relative to semantic
or structural edits.  Figure~\ref{fig:taxonomy} shows the type hierarchy slice
used for the schematic examples.

\begin{figure}[t]
\centering
\begin{tikzpicture}[x=1cm,y=1cm]
  \node[trnd-root, minimum width=2.2cm] (root) at (6.25,0) {registry root};

  \node[trnd-trig] (trigger) at (1.55,-1.35) {trigger};
  \node[trnd-out] (output) at (5.25,-1.35) {output};
  \node[trnd-inp] (transform) at (8.00,-1.35) {transform};
  \node[trnd-lc] (lang) at (10.90,-1.35) {langchain / llms};
  \foreach \x in {trigger,output,transform,lang}{\draw[tree-e] (root) -- (\x);}

  \node[trnd-tp-trig] (webhook) at (0.35,-2.75) {Webhook};
  \node[trnd-tp-trig] (manual) at (2.75,-2.75) {Manual Trigger};
  \draw[tree-e] (trigger) -- (webhook);
  \draw[tree-e] (trigger) -- (manual);

  \node[trnd-tp-out] (http) at (5.25,-2.75) {HTTP Request};
  \draw[tree-e] (output) -- (http);

  \node[trnd-tp-inp] (set) at (8.00,-2.75) {Edit Fields};
  \draw[tree-e] (transform) -- (set);

  \node[trnd-lc, minimum width=2.45cm] (llmgroup) at (10.90,-2.65) {chat models};
  \node[trnd-tp-lc] (ollama) at (9.65,-4.00) {Ollama Chat};
  \node[trnd-tp-lc] (openai) at (12.15,-4.00) {OpenAI Chat};
  \draw[tree-e] (lang) -- (llmgroup);
  \draw[tree-e] (llmgroup) -- (ollama);
  \draw[tree-e] (llmgroup) -- (openai);

  \node[font=\scriptsize, text=graydark!82, align=left, anchor=west] at (-0.15,-4.65)
    {Pinned snapshot: maximum depth $3$, diameter $6$.\\
     Taxonomic distance is the shortest-path length divided by the fixed diameter.};
\end{tikzpicture}
\caption{Illustrative fragment of the pinned n8n type hierarchy.  Input-oriented
triggers, output-oriented HTTP requests, transformations, and language-model
nodes occupy distinct branches.  The complete snapshot contains additional
branches; 72 of the 469 corpus types were absent from the pinned registry and
were attached to the root, a limitation discussed in Section~\ref{sec:discussion}.}
\Description{A hierarchy fragment places trigger, output, transformation, and language-model node types under a common registry root and illustrates shortest-path taxonomic distance.}
\label{fig:taxonomy}
\end{figure}

\subsection{Field distance}

Pair-dependent field sets or scales can break the triangle inequality.  We
therefore fix a nonempty finite field universe $\K_\star$.  Each field has a
tagged canonicalizer $c_k$ into a fixed domain $\mathcal X_k$ containing an
out-of-band missing element $\missing_k$, a metric
$\delta_k:\mathcal X_k^2\to[0,1]$, and a weight $\omega_k>0$.  A canonical
dictionary contains exactly one $c_k$ value for every $k\in\K_\star$.  Then
\begin{equation}\label{eq:field_distance}
\dfield(a,b)=
\frac{\sum_{k\in\K_\star}\omega_k\delta_k(a[k],b[k])}
     {\sum_{k\in\K_\star}\omega_k}
\end{equation}
is a metric on canonical dictionaries.

\begin{table}[t]
\centering
\caption{Typed field primitives used by the metric implementation.  Every
primitive is evaluated on a fixed canonical domain and contributes to the same
pair-independent field universe.}
\label{tab:field_primitives}
\begin{tabularx}{\textwidth}{@{} l Y Y @{} }
\toprule
Field class & Canonical representation & Metric primitive \\
\midrule
Categorical & tagged categorical value & discrete distance \\
Boolean & canonical Boolean value & discrete distance \\
Numeric & canonical finite scalar & $\min\{1,|a-b|/R_k\}$ \\
String or expression & canonical string serialization &
$\min\{1,\mathrm{Lev}(a,b)/L_k\}$ \\
JSON object or list & deterministic type-preserving serialization &
$\min\{1,\mathrm{Lev}(a,b)/L_k\}$ \\
Missing & reserved field-specific element $\missing_k$ & declared domain metric \\
\bottomrule
\end{tabularx}
\end{table}

Clipping a metric at a fixed positive threshold preserves the triangle
inequality.  The implementation applies the fixed canonical domains and
primitive distances in Table~\ref{tab:field_primitives} over a frozen universe
of 2,930 field keys.  Equal positive weights provide a transparent,
pair-independent baseline in which every registered field can contribute and no
pair is renormalized according to the fields it happens to contain.  A Level~1
change therefore produces the expected local contribution to the complete
workflow distance, while larger node and topology changes accumulate across
multiple terms.

\subsection{Substitution and null edits}

For $w_\tau,w_f>0$ with $w_\tau+w_f=1$, define
\begin{equation}\label{eq:node_distance}
\dnode(u,v)=w_\tau\dtype(\tau(u),\tau(v))
            +w_f\dfield(\alpha(u),\alpha(v)).
\end{equation}
It is a metric on canonical attribute states $(\tau(v),\alpha(v))$ and is bounded
by one; node identities themselves are deliberately irrelevant.  Let
$\mathcal R_0$ be the fixed set containing one empty default attribute state for every
type, and set
\begin{equation}
\kappa(v)=\min_{r\in\mathcal R_0}\dnode(v,r),\qquad
\mu(v)=\beta+\gamma\kappa(v),
\label{eq:null_cost}
\end{equation}
with $\beta\ge1/2$ and $0\le\gamma\le1$.  Distance to a fixed set is
1-Lipschitz, so $|\kappa(u)-\kappa(v)|\le\dnode(u,v)$.  Hence the extension
\begin{equation}
\dnodebar(u,v)=\dnode(u,v),\quad
\dnodebar(u,\varepsilon)=\dnodebar(\varepsilon,u)=\mu(u),\quad
\dnodebar(\varepsilon,\varepsilon)=0
\end{equation}
is a metric: $\dnode(u,v)\le1\le2\beta\le\mu(u)+\mu(v)$, and the
Lipschitz bound gives the other two triangle orientations through
$\varepsilon$.  The evaluated default is
$w_\tau=w_f=\beta=\gamma=1/2$.

\section{Formal exact semantic graph matching}\label{sec:metric}

\begin{definition}[Partial matching]
A partial node matching $M\subseteq V_1\times V_2$ contains each node at most
once.  Its unmatched sets are $V_1\setminus\dom(M)$ and
$V_2\setminus\im(M)$.
\end{definition}

Given $M$, identify every matched pair as one representative and retain every
unmatched node as a graph-tagged singleton.  This produces a common universe
$U_M$ and maps $\sigma_i^M:V_i\to U_M$.  Lift the edge sets as
\begin{equation}
\widetilde E_i^M=\{(\sigma_i^M(a),\sigma_i^M(b),\ell):(a,b,\ell)\in E_i\}.
\end{equation}
The node and edge costs induced by $M$ are
\begin{align}
C_N(M)&=\sum_{(u,v)\in M}\dnode(u,v)
 +\sum_{u\notin\dom(M)}\mu(u)+\sum_{v\notin\im(M)}\mu(v),\\
C_E(M)&=|\widetilde E_1^M\triangle\widetilde E_2^M|.
\end{align}

\begin{definition}[Workflow distance]\label{def:workflow_distance}
For $w_n,w_e>0$, the exact workflow distance is
\begin{equation}\label{eq:workflow_distance}
\dw(G_1,G_2)=\min_{M\in\M(V_1,V_2)}
 \{w_nC_N(M)+w_eC_E(M)\}.
\end{equation}
\end{definition}

\paragraph{Relation to graph edit distance.}
Equation~\eqref{eq:workflow_distance} is a domain-specific node-map graph
edit distance: the partial correspondence induces node substitutions, node
insertions and deletions, and the remaining directed labelled-edge edits.
The general correspondence-based construction has established antecedents
\cite{justice2006binary,bougleux2017qap}.  The specific contribution here is
the workflow representation and metric-compatible type, field, null, and
edge cost contract, together with the proof and controlled validation of the
resulting workflow metric.

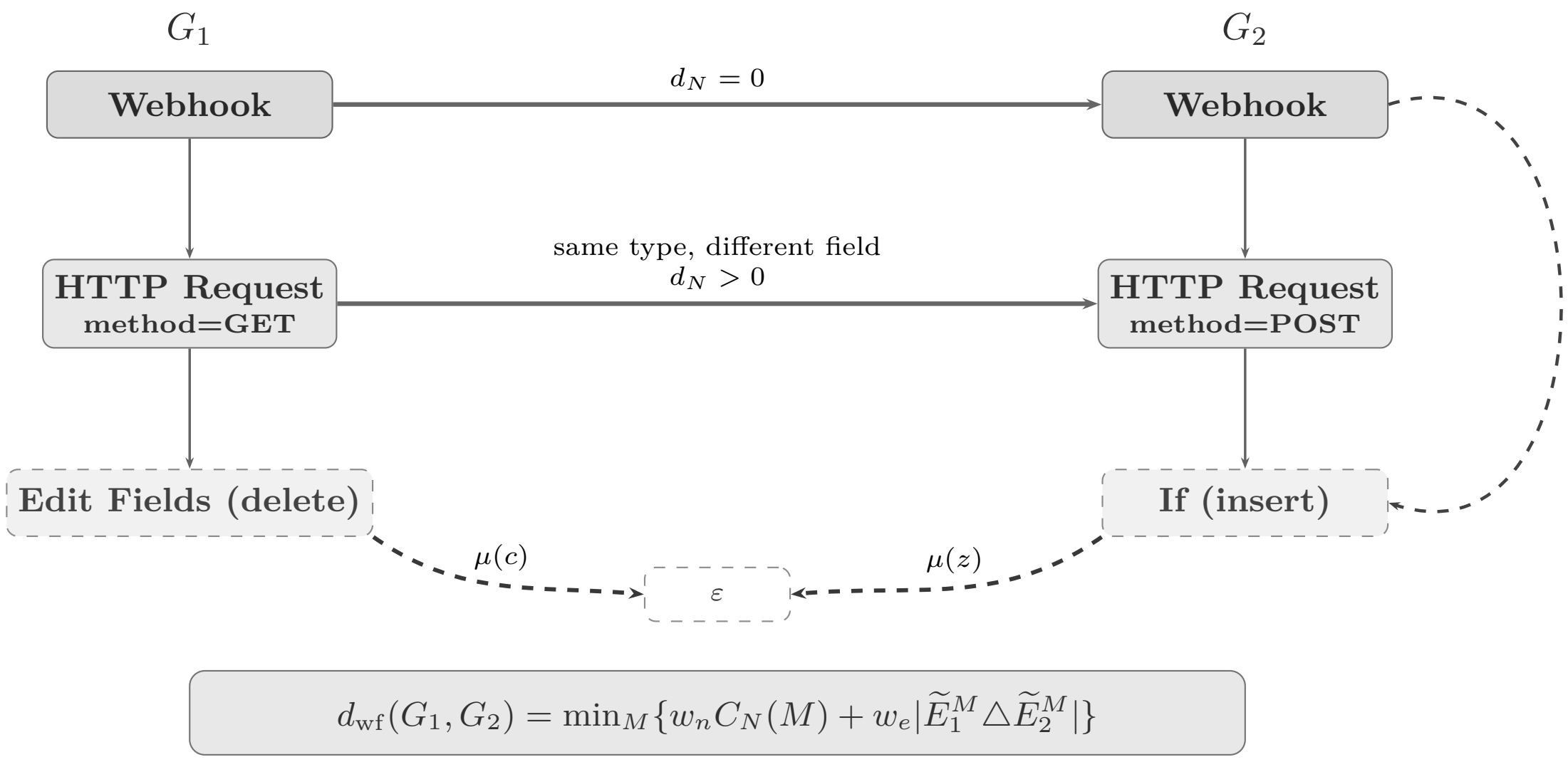
\begin{figure}[t]
\centering
\begin{tikzpicture}[x=1cm,y=1cm]
  \fill[graylight!70, rounded corners=6pt] (-1.8,1.05) rectangle (2.0,-4.8);
  \fill[graylight!70, rounded corners=6pt] (8.0,1.05) rectangle (11.8,-4.8);
  \node[font=\bfseries, text=graydark] at (0.1,0.70) {$G_1$};
  \node[font=\bfseries, text=graydark] at (9.9,0.70) {$G_2$};

  \node[nd-trig] (a) at (0.1,0) {Webhook};
  \node[nd-out, minimum height=0.82cm] (b) at (0.1,-1.85)
    {HTTP Request\\[-2pt]\scriptsize method=GET};
  \node[nd-ins] (c) at (0.1,-3.70) {Edit Fields (delete)};
  \draw[wfedge] (a) -- (b);
  \draw[wfedge] (b) -- (c);

  \node[nd-trig] (x) at (9.9,0) {Webhook};
  \node[nd-out, minimum height=0.82cm] (y) at (9.9,-1.85)
    {HTTP Request\\[-2pt]\scriptsize method=POST};
  \node[nd-ins] (z) at (9.9,-3.70) {If (insert)};
  \draw[wfedge] (x) -- (y);
  \draw[wfedge] (y) -- (z);
  \draw[wfedge-disc] (x.east) to[out=18,in=-18,looseness=1.55] (z.east);

  \draw[arr-id] (a.east) -- node[above, font=\scriptsize] {$d_N=0$} (x.west);
  \draw[arr-id] (b.east) -- node[above, font=\scriptsize, align=center]
    {same type, different field\\$d_N>0$} (y.west);
  \node[trnd-root, dashed, minimum width=1.35cm] (eps) at (5.0,-4.55) {$\varepsilon$};
  \draw[arr-sub] (c.south east) to[out=-35,in=175]
    node[above, font=\scriptsize] {$\mu(c)$} (eps.west);
  \draw[arr-sub] (z.south west) to[out=-145,in=5]
    node[above, font=\scriptsize] {$\mu(z)$} (eps.east);

  \node[draw, rounded corners=4pt, fill=navylight, draw=navymid,
    text=navydark, align=center, minimum width=9.8cm, minimum height=0.78cm,
    font=\small] at (5.0,-5.65)
    {$d_{\mathrm{wf}}(G_1,G_2)=\min_M\{w_n C_N(M)+w_e
      |\widetilde E_1^M\triangle\widetilde E_2^M|\}$};
\end{tikzpicture}
\caption{One candidate partial matching.  The two HTTP Request nodes share a
type but differ in their declared method, so only their field component changes.
Unmatched nodes pay null-edit costs, and the burgundy dashed connection exists
only in $G_2$.  Node and edge costs are evaluated under the same matching.}
\Description{Two small workflow graphs are aligned. Matched HTTP Request nodes differ in a parameter, unmatched nodes map to the null symbol, and one directed edge exists only in the second graph.}
\label{fig:metric_breakdown}
\end{figure}

Figure~\ref{fig:metric_breakdown} decomposes the two cost channels in
Definition~\ref{def:workflow_distance}.

The empty matching supplies the pair-dependent upper bound
\begin{equation}\label{eq:normalizer}
Z(G_1,G_2)=w_n\!\left(\sum_{u\in V_1}\mu(u)+
\sum_{v\in V_2}\mu(v)\right)+w_e(|E_1|+|E_2|).
\end{equation}
When $Z>0$, $1-\dw/Z\in[0,1]$ is a reporting similarity.  It is not the
metric, and it was not persisted in the canonical result files; all empirical
plots below therefore report raw distance.

\begin{assumption}[Frozen metric context]\label{ass:frozen}
The type set, hierarchy and its diameter, exactly normalized embedding table,
edge-label domain and canonicalizer, nonempty field universe, typed field
canonicalizers and out-of-band missing elements, field scales, primitive field
metrics, and all weights are frozen before any comparison.  Type and node
weights are positive where required, the three type weights sum to one with
$\lambda_0>0$, $\beta\ge1/2$, $0\le\gamma\le1$, and $w_n,w_e>0$.
\end{assumption}

\begin{theorem}[Metric validity]\label{thm:metric}
Under Assumption~\ref{ass:frozen}, $\dw$ is a metric on finite canonical
attributed directed labelled simple graphs, considered up to attributed
directed labelled graph isomorphism.
\end{theorem}

Appendix~\ref{app:proof} proves the theorem.  The triangle argument composes two
partial matchings, embeds all three graphs in the equivalence closure induced by
those matchings, and uses contraction of edge-set symmetric difference under
quotient maps.  This last step is necessary; treating pairwise quotient spaces
as if they were already identical would leave a gap.

\section{Approximate matching}\label{sec:approximation}

Exact edge-aware matching is combinatorial and factorial-scale in the balanced
worst case.  The implementation
uses branch and bound, seeded by a feasible assignment and pruned with node and
forced-edge lower bounds (Algorithm~\ref{alg:compute_dw} in
Appendix~\ref{app:computation}).  The retained code prunes and updates with a
$10^{-12}$ tolerance; a completed run is therefore solver-complete to that
numerical tolerance, not a certificate of an exact real-valued optimum.

For larger pairs, a padded linear assignment \cite{kuhn1955hungarian} selects a matching from node
substitution and null costs; that matching is then scored with the full node and
edge objective.  We denote the result $\dwh$.  Because it is one feasible
matching for Equation~\eqref{eq:workflow_distance}, $\dwh\ge\dw$ whenever both
use the same frozen context.  Selection ignores edges, so $\dwh$ need not retain
the metric axioms.  With $N=|V_1|+|V_2|$, the assignment step is $O(N^3)$ after
the pairwise ground costs have been constructed.

The full-results artifact records whether each evaluation completed
branch-and-bound search.  The row-level \texttt{used\_exact} field therefore
identifies the optimizer used, not a different distance definition: every row is
scored with the node-and-edge objective in
Equation~\eqref{eq:workflow_distance}.  Throughout the empirical analysis, a
\emph{branch-and-bound-complete} evaluation is one for which search terminates
for the implemented finite-precision objective under the declared
$10^{-12}$ tolerance.  This denotes optimization completeness rather than exact
real arithmetic.  Main inferential statements use only these rows and complete
hash-level panels; assignment-based upper bounds are never pooled with them.

\section{Corpus and artifact provenance}\label{sec:corpus}

\subsection{Corpus snapshot}

The corpus builder scanned JSON files from three public GitHub repositories; it
did not ingest the official n8n template API.  The locally recoverable source
commits are listed in Appendix~\ref{app:provenance}.  The resulting snapshot
contains 4,705 accepted path-identified workflow graphs but only 2,282 distinct
persisted normalized-graph hashes.  Table~\ref{tab:corpus_profile} reports the
corpus quantities actually consumed by the experiments.

\begin{table}[htbp]
\centering
\caption{Canonical corpus profile reconstructed from
\texttt{dataset\_full/corpus}.}
\label{tab:corpus_profile}
\begin{tabularx}{\textwidth}{@{} l r Y @{} }
\toprule
Quantity & Value & Interpretation \\
\midrule
Path-identified workflow graphs & 4,705 & Accepted public JSON exports after parser checks \\
Distinct persisted normalized-graph hashes & 2,282 & Duplicate-aware unit used in the main panels \\
Node rows & 73,057 & Median 12; IQR 6--20; maximum 246 nodes per workflow \\
Observed node types & 469 & Broad coverage of base and agentic workflow components \\
Canonical field keys & 2,930 & Frozen pair-independent field universe and scale table \\
Most frequent type & 17,308 rows & \texttt{n8n-nodes-base.stickyNote} \\
Source repositories & 3 & Public collections pinned to the commits in Appendix~\ref{app:provenance} \\
\bottomrule
\end{tabularx}
\end{table}

The pinned hierarchy records \texttt{n8n-nodes-base} 2.15.1 and
\texttt{@n8n/n8n-nodes-langchain} 2.16.1.  Seventy-two observed types not present
in that registry snapshot receive a common-root attachment, yielding complete
coverage without pair-specific hierarchy changes.  The frozen description
encoder is \path{sentence-transformers/all-MiniLM-L6-v2}
\cite{allminilmmodelcard}; its embedding table is fixed for every comparison.

\subsection{Recovered lineage and remaining gaps}

The corpus ID is the first 12 hexadecimal characters of a SHA-1 hash of the
absolute source \emph{path}, not file content.  The persisted tables discard that
path, but the exact Windows checkout layout survives.  Rehashing it reproduces
all 4,705 IDs with no observed collision.  The new
\texttt{recovered\_source\_manifest.csv} records repository-relative paths,
source commits, file and candidate hashes, and a persisted normalized-graph hash.
This reconstruction establishes a one-to-one lineage from every evaluated ID
to a repository-relative source path and content hash.  Subsequent analyses use
the content-derived hash rather than the path-derived ID whenever duplicate
control is required.

The recovery exposes extensive copying across the three collections.  Only
2,282 normalized-graph hashes remain after grouping; 4,671 IDs occur in cross-source
duplicate groups, with a largest group of 20.  Moreover, 2,248 of 2,282 hash groups appear in more than one source.  The
repositories are therefore treated as overlapping collections, and all main
summaries equal-weight distinct persisted normalized-graph hashes rather than source
paths.

The parser retains parameter values because they contribute to semantic
comparison.  The released paper reports aggregate results and source manifests;
redistribution of raw exports remains governed by the corresponding repository
licences.

\section{Empirical validation}\label{sec:empirical}

The empirical study evaluates construct validity of the proposed metric through
five complementary tests on real workflow graphs: invariance to
representation-preserving changes, sensitivity to graded semantic and structural
changes, accumulation under repeated edits, discrimination between systems with
different perturbation burdens, and robustness to metric weighting.  A separate
triangle audit checks numerical consistency with the formal result.  All
reported tests evaluate the metric in
Equation~\eqref{eq:workflow_distance} under the same frozen context; the
stored solver flag records whether its minimum was completed by branch and
bound.  The empirical objective is internal construct validation rather than
a claim of superiority over every workflow-similarity measure.  The tests ask
whether the proposed metric exhibits the stability, directional sensitivity,
accumulation, weighting robustness, and numerical consistency implied by its
construction.  Table~\ref{tab:perturbations} records the transformation
library.

\begin{table}[t]
\centering
\caption{Controlled transformation library used in the empirical validation.}
\label{tab:perturbations}
\begin{tabularx}{\textwidth}{@{} c l Y Y @{} }
\toprule
Level & Operators & Transformation class & Principal metric terms \\
\midrule
L0 & \texttt{rename\_node}, \texttt{shuffle\_order}
& Representation-preserving surface change
& Near-zero finite-precision residual \\
L1 & \texttt{toggle\_bool}, \texttt{change\_str}, \texttt{change\_cat}
& Single parameter-value change
& Field substitution \\
L2 & \texttt{insert\_node}, \texttt{delete\_node}, \texttt{rewire\_edge}
& Local node or routing change
& Null edits and edge disagreement \\
L3 & \texttt{swap\_type}, \texttt{merge\_workflow}, \texttt{replace\_subgraph}
& Major type or multi-node structural change
& Type, null, and edge terms \\
\bottomrule
\end{tabularx}
\end{table}

\paragraph{Protocol.}
Experiments~1 and~2 use branch-and-bound-complete rows marked
\texttt{used\_exact=true}.  Complete hash-level panels are formed before
summarization so that each distinct workflow contributes equally at every level
or chain length.  Branch-and-bound-complete and assignment-based evaluations are kept separate
throughout.

\begin{figure}[t]
\centering
\includegraphics[width=\textwidth]{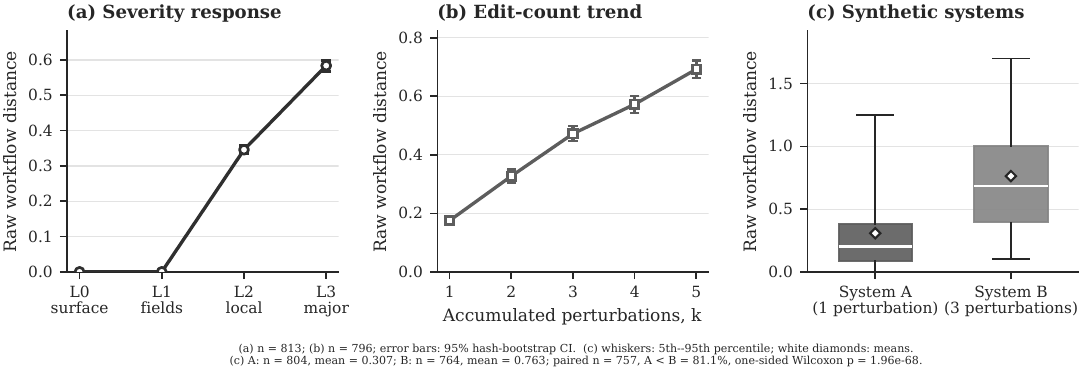}
\caption{Core branch-and-bound-complete results after grouping source-path copies by
persisted normalized-graph hash.  Panels (a) and (b) show equal-hash means and
95\% hash-bootstrap intervals on complete panels.  Panel (c) shows marginal
hash-level distributions with 5th--95th-percentile whiskers; its paired
comparison uses the 757 hashes present in both systems.}
\Description{Three panels show that mean workflow distance increases across surface, field, local, and major transformations; increases with one to five accumulated perturbations; and separates systems with one versus three perturbations.}
\label{fig:validation_core}
\end{figure}

Experiment~1 applies five independently generated transformations per level.
Figure~\ref{fig:validation_core} equal-weights the 813 normalized-graph hashes with an
exact value at every level and uses hash-level bootstrap intervals.
Experiment~2 applies chains of one to five Level~1 or Level~2 operations and
measures rank association on a complete 796-hash panel.  Experiment~3 varies the
metric weights and is reported in Appendix~\ref{app:experiments}.
Experiment~4 compares systems receiving one and three perturbations.
Experiment~5 evaluates stored triangle slacks as a numerical consistency check
of the metric implementation.

\paragraph{Graded transformation response.}
Across 32,093 branch-and-bound-complete Experiment~1 evaluations, the complete-panel equal-hash
means are $1.21\times10^{-5}$ at L0, $7.84\times10^{-5}$ at L1, $0.346$ at
L2, and $0.584$ at L3.  The L0 value is a numerically negligible finite-precision
floor and is more than four orders of magnitude below the major-change mean.
Hash-level transitions are nondecreasing for 100.0\%, 99.75\%, and 84.01\%
of the 813 hashes from L0$\to$L1, L1$\to$L2, and L2$\to$L3.  The first two
transitions are essentially universal; the remaining overlap is expected
because L2 and L3 denote heterogeneous operator families rather than fixed
single-cost increments.  The aggregate response nevertheless exhibits the
intended ordering from surface stability through local change to major semantic
or structural modification.

\paragraph{Accumulated edits.}
Experiment~2 contains 8,031 branch-and-bound-complete evaluations.  On the 796 normalized-graph hashes
with all five steps, equal-hash mean distance increases from $0.175$ at $k=1$ to
$0.692$ at $k=5$, and the pooled rank association is $\rho_S=0.520$.  The
within-hash median correlation is $0.975$, with 96.4\% of workflows showing a
positive association.  The difference between the pooled and within-workflow
coefficients reflects genuine heterogeneity in graph size and edit composition;
within each workflow, the metric tracks accumulated change strongly.

\paragraph{Synthetic system discrimination.}
Equal-weighting normalized-graph hashes gives marginal means $0.307$ for System~A
(804 hashes) and $0.763$ for System~B (764).  On 757 paired hashes, the means are
$0.309$ and $0.761$; System~A has lower distance for 81.1\% of workflows, with
a one-sided paired Wilcoxon $p=1.96\times10^{-68}$.  The metric therefore
separates two systems that differ only in their controlled perturbation burden,
showing that the aggregate score remains discriminative beyond individual edit
operators.

\paragraph{Weight and triangle audits.}
All five positive-weight configurations retain a positive edit-count
association, with $\rho_S$ from 0.507 to 0.584 on the common hash panel; their
different raw scales and node/edge shares are documented in
Appendix~\ref{app:experiments}.  The triangle analysis contains 117,312 ordered orientations of 19,552
unordered graph triples from 50 graph IDs and 49 normalized-graph hashes.  No recorded
slack is negative and the minimum is $9.1585\times10^{-6}$.  The formal distance is zero for identical canonical graphs.  The observed
same-content residual of $1.8317\times10^{-5}$ arises from finite-precision
evaluation of the embedding term and is reported as an implementation-level
numerical error.  It is of the same order as the L0 residual and several orders
of magnitude below the distances induced by substantive graph changes.  The
audit therefore provides broad numerical support
for the triangle property established analytically in
Theorem~\ref{thm:metric}.

\section{Discussion}\label{sec:discussion}

The central contribution is a metric measurement layer for workflow artifacts.
It makes representation, node correspondence, parameter disagreement, null
edits, topology disagreement, and solver status inspectable rather than
compressing them into an opaque similarity score.  A workflow can be different
because the same operation is configured differently, because an operation is
missing, or because familiar operations are connected through a different
route.  The shared partial matching places those possibilities in one coherent
objective and preserves a component-level audit trail.

The empirical results support the intended construct from several independent
directions.  First, representation-preserving changes remain at a numerical
zero floor.  Second, mean distance increases from surface and field changes to
local and major graph modifications, with nearly universal monotonicity through
the first two transitions.  Third, repeated edits accumulate consistently:
96.4\% of workflows have a positive edit-count association and the median
within-workflow correlation is 0.975.  Fourth, the metric separates one- and
three-perturbation systems by a large margin.  Fifth, the positive association
persists across node-heavy, edge-heavy, field-dominant, and type-and-edge
weightings, showing that the result is not an artefact of one balanced
configuration.  Finally, 117,312 tested triangle orientations show no negative
slack.  Taken together, these tests support the intended numerical and directional
properties of the metric: stability under surface-preserving transformations,
sensitivity to parameter and graph changes, accumulation under repeated edits,
discrimination between perturbation burdens, robustness to weighting choices,
and consistency with the analytically established triangle property.

The distinction between the exact metric and its Hungarian-first upper bound is
important in practice.  Exact edge-aware matching provides the reference value
and the metric guarantee; the assignment procedure provides a feasible matching
with cubic assignment cost after ground distances are constructed.  Because the
score decomposes into node and edge contributions, both procedures can support
workflow-generator regression testing, near-duplicate control, template
retrieval, migration auditing, and maintenance analysis while still showing
which part of the artifact changed.

The metric is deliberately scoped to the represented workflow artifact.
Credentials, external state, service versions, and runtime data can make
structurally identical workflows behave differently, while distinct workflows
can implement the same external function.  The corpus comprises public n8n
collections rather than a platform-usage sample, and exact matching remains
combinatorial for large pairs.  Runtime validation and learned approximators address complementary questions;
the formal result and controlled construct validation reported here remain
well defined at the artifact level.

\section{Conclusion}\label{sec:conclusion}

This paper introduced an exact, auditable metric for comparing no-code
workflows as attributed directed labelled graphs.  The distance evaluates type,
parameter, null-edit, and routing disagreement under one partial node
correspondence and is a metric under a frozen comparison context.  A
Hungarian-first procedure provides a clearly identified feasible upper bound for
larger comparisons.

The evaluation covers 4,705 public workflows and 32,093 exact graded-change
comparisons.  Surface-preserving transformations remain at a numerical zero
floor, major type or subgraph changes reach a mean distance of 0.584, accumulated
edit count has pooled Spearman $\rho=0.520$ and median within-workflow
correlation 0.975, and one- versus three-perturbation systems separate at 0.307
versus 0.763.  The association is stable across five weight configurations, and
117,312 triangle orientations show no negative slack.  These results support the
metric as a reproducible artifact-level instrument for workflow evaluation,
retrieval, duplicate control, migration analysis, and regression testing.

\appendix

\section{Proof of metric validity}\label{app:proof}

This appendix proves Theorem~\ref{thm:metric}.  All type, field, and graph
objects below use the same frozen context $\Theta$ from
Assumption~\ref{ass:frozen}.

\begin{lemma}[Ground metrics]\label{lem:ground_metrics}
The type distance $\dtype$, field distance $\dfield$, node distance $\dnode$,
and null-extended node distance $\dnodebar$ are metrics on their respective
canonical domains, and $0\le\dnode\le1$.
\end{lemma}

\begin{proof}
Shortest-path distance on a fixed hierarchy is a metric.  Angular distance is a
metric on normalized vectors and can only become a pseudometric after mapping
type identifiers to vectors.  The discrete identity term is a metric and has
positive coefficient $\lambda_0$, so Equation~\eqref{eq:type_distance} is a
metric on type identifiers.  Its three coefficients sum to one, hence it is
bounded by one.

Each $\delta_k$ is a metric on its fixed tagged field domain including the
out-of-band missing element.  Equation~\eqref{eq:field_distance} is a positive
normalized sum of those metrics, hence is a metric bounded by one.  The same
argument applies to $\dnode$ on canonical attribute states because
$w_\tau,w_f>0$ and sum to one.  Table~\ref{tab:field_primitives} records the
fixed canonical domains and primitives used by the implementation.

It remains to check triangles through $\varepsilon$.  Since distance to a
nonempty fixed set is 1-Lipschitz,
$|\kappa(u)-\kappa(v)|\le\dnode(u,v)$.  Thus
\begin{align}
\dnodebar(u,v)&=\dnode(u,v)\le1\le2\beta
 \le\mu(u)+\mu(v),\\
\mu(u)-\mu(v)&=\gamma(\kappa(u)-\kappa(v))
 \le\gamma|\kappa(u)-\kappa(v)|\le\dnode(u,v),
\end{align}
and symmetrically $\mu(v)-\mu(u)\le\dnode(u,v)$.  These are the three
triangle orientations containing $\varepsilon$.  Positivity of $\beta$ gives
identity of indiscernibles at the null symbol.
\end{proof}

\begin{lemma}[Symmetric difference contracts under a map]\label{lem:contraction}
For finite sets $A,B$ and a map $q$ on their elements,
\begin{equation}
q(A)\triangle q(B)\subseteq q(A\triangle B),\qquad
|q(A)\triangle q(B)|\le|A\triangle B|.
\end{equation}
\end{lemma}

\begin{proof}
If $x\in q(A)\setminus q(B)$, some preimage of $x$ is in $A$ and no preimage
of $x$ is in $B$; therefore a preimage lies in $A\setminus B$.  The reverse
orientation is identical.  Taking cardinalities gives the second statement.
\end{proof}

\begin{lemma}[Composition bound]\label{lem:composition}
Let $M_{12}$ match $G_1$ to $G_2$ and $M_{23}$ match $G_2$ to $G_3$.  Define
their partial composition
\begin{equation}
M_{13}=\{(u,w):\exists v\in V_2, (u,v)\in M_{12}, (v,w)\in M_{23}\}.
\end{equation}
Then
\begin{align}
C_N(M_{13})&\le C_N(M_{12})+C_N(M_{23}),\label{eq:node_comp_bound}\\
C_E(M_{13})&\le C_E(M_{12})+C_E(M_{23}).\label{eq:edge_comp_bound}
\end{align}
\end{lemma}

\begin{proof}
Draw the matched pairs as links in the disjoint union
$V_1\sqcup V_2\sqcup V_3$.  Each connected component contains at most one
node from each graph.  For a three-node component $(u,v,w)$, the node
contribution satisfies
$\dnodebar(u,w)\le\dnodebar(u,v)+\dnodebar(v,w)$.  For an $(u,v)$ component
with no $w$, it satisfies
$\mu(u)\le\dnode(u,v)+\mu(v)$; a $(v,w)$ component is symmetric.  A singleton
from $V_1$ or $V_3$ has the same null cost on the corresponding side, while a
singleton from $V_2$ contributes two nonnegative null costs on the right and
nothing on the left.  Summing over components proves
Equation~\eqref{eq:node_comp_bound}.

For edges, let $\sim$ be the equivalence closure of
$M_{12}\cup M_{23}$ on the same disjoint union, let $Q$ be its quotient, and
let $F_i$ be the image of $E_i$ in the common directed labelled edge universe
$Q\times Q\times\Lambda$.  The natural map
$r_{13}:U_{M_{13}}\to Q$ is injective on the classes meeting
$V_1\cup V_3$: two such classes merge in $Q$ only through a common node of
$V_2$, which is exactly the relation used to form $M_{13}$.  Its induced map on
labelled edges is therefore injective.  Consequently
\begin{equation}
C_E(M_{13})=|F_1\triangle F_3|.
\end{equation}
Let $q_{12}:U_{M_{12}}\to Q$ and $q_{23}:U_{M_{23}}\to Q$ be the natural maps;
neither needs to be surjective.  For $i\in\{1,2\}$,
$F_i=q_{12}(\widetilde E_i^{M_{12}})$, and for $i\in\{2,3\}$,
$F_i=q_{23}(\widetilde E_i^{M_{23}})$.  Lemma~\ref{lem:contraction} therefore
gives $|F_1\triangle F_2|\le C_E(M_{12})$ and
$|F_2\triangle F_3|\le C_E(M_{23})$.  The Hamming triangle inequality for
finite sets gives
\begin{equation}
|F_1\triangle F_3|
\le |F_1\triangle F_2|+|F_2\triangle F_3|
\le C_E(M_{12})+C_E(M_{23}),
\end{equation}
which proves Equation~\eqref{eq:edge_comp_bound}.
\end{proof}

\begin{proof}[Proof of Theorem~\ref{thm:metric}]
Nonnegativity follows from the component costs.  Inverting a partial matching
bijection preserves its node cost and exchanges the two lifted edge sets, so
symmetry follows.  The set of partial matchings is finite, hence the minimum
exists.  Attributed isomorphisms transport partial matchings bijectively and
preserve both component costs; the value is therefore independent of the chosen
representatives of two isomorphism classes.

If $G_1$ and $G_2$ are attributed directed labelled isomorphic, their
isomorphism is a full matching with zero node and edge cost.  Conversely, if
$\dw(G_1,G_2)=0$, positivity of $w_n,w_e$ forces both component costs to zero.
Since $\mu(v)>0$, no node is unmatched; the matching is a bijection.  Since
$\dnode$ is a metric, matched node attributes agree.  Zero edge symmetric
difference says that the same bijection preserves every directed labelled edge.
It is therefore an attributed graph isomorphism.

For the triangle inequality, choose optimal matchings $M_{12}$ and $M_{23}$ and
compose them as in Lemma~\ref{lem:composition}.  Since $M_{13}$ is feasible,
\begin{align}
\dw(G_1,G_3)
&\le w_nC_N(M_{13})+w_eC_E(M_{13})\\
&\le w_n\{C_N(M_{12})+C_N(M_{23})\}
   +w_e\{C_E(M_{12})+C_E(M_{23})\}\\
&=\dw(G_1,G_2)+\dw(G_2,G_3).
\end{align}
Thus all four metric axioms hold and the distance descends to attributed graph
isomorphism classes.
\end{proof}

\section{Formal exact and numerical approximate computation}\label{app:computation}

\begin{algorithm}[t]
\caption{Branch and bound for the formal edge-aware objective}
\label{alg:compute_dw}
\begin{algorithmic}[1]
\Require Canonical graphs $G_1,G_2$; frozen context $\Theta$; tolerance
$\epsilon_{\mathrm{BB}}\ge0$
\Ensure A matching $M^\star$ and incumbent upper bound $U$
\State Compute all substitution costs $d_N(u,v)$ and null costs $\mu(u),\mu(v)$
\State $M^\star\gets$ padded Hungarian node assignment
\State $U\gets w_nC_N(M^\star)+w_eC_E(M^\star)$ \Comment{feasible upper bound}
\State Put the nodes of $V_1$ in a fixed search order
\State Initialize the stack with the empty partial function
$\phi:V_1\rightharpoonup V_2\cup\{\varepsilon\}$
\While{the stack is not empty}
  \State Remove one state $\phi$ from the stack
  \State $L_N\gets C_N^{\mathrm{committed}}(\phi)+L_N^{\mathrm{LAP}}(\phi)$
  \State $L_E\gets$ edge disagreements already forced by $\phi$
  \If{$w_nL_N+w_eL_E\ge U-\epsilon_{\mathrm{BB}}$}
    \State \textbf{continue} \Comment{no improvement beyond the tolerance}
  \EndIf
  \If{$\phi$ assigns every node of $V_1$}
    \State Form $M_\phi=\{(u,v):\phi(u)=v\in V_2\}$ and evaluate its full cost
    \If{$w_nC_N(M_\phi)+w_eC_E(M_\phi)<U-\epsilon_{\mathrm{BB}}$}
      \State $(M^\star,U)\gets(M_\phi,w_nC_N(M_\phi)+w_eC_E(M_\phi))$
    \EndIf
  \Else
    \State Choose next $u\in V_1$ and extend $\phi(u)=v$ for each unused $v\in V_2$
    \State Push the deletion branch $\phi(u)=\varepsilon$
  \EndIf
\EndWhile
\State \Return $(M^\star,U)$
\end{algorithmic}
\end{algorithm}

The initial assignment supplies a valid upper bound; it does not determine the
formal optimum.  With conservative lower bounds, exhaustion returns a value
within the stated numerical tolerance; $\epsilon_{\mathrm{BB}}=0$ gives the ideal
exact algorithm.  The number of complete partial injections is
$\sum_{k=0}^{\min(n_1,n_2)}\binom{n_1}{k}n_2!/(n_2-k)!$, factorial-scale when
$n_1=n_2$.  The assignment bound costs $O(N^3)$ per uncached subproblem, where
$N=n_1+n_2$; that polynomial subroutine does not make the search polynomial.

The branch-and-bound implementation evaluates a complete state only after all
nodes of $V_1$ have been assigned to a distinct node of $V_2$ or to the null
symbol.  Its node lower bound adds the committed cost to a padded assignment on
the remainder; its edge bound counts disagreements already forced by committed
endpoints.  Ignoring edge labels in that incremental edge bound is conservative:
it can weaken pruning but cannot prune an optimum.  At a leaf, the full directed
labelled symmetric difference is evaluated.  The stored code compares bounds and
incumbents with $\epsilon_{\mathrm{BB}}=10^{-12}$, so a completed search is an
additive-tolerance numerical result.  Setting $\epsilon_{\mathrm{BB}}=0$ in exact
arithmetic describes the ideal formal solver.

Table~\ref{tab:parameters} records the empirical parameters used by the main
weighted distance outside the ablation.

\begin{table}[t]
\centering
\caption{Balanced metric parameters used outside the weight ablation.}
\label{tab:parameters}
\begin{tabularx}{\textwidth}{@{} l l Y @{} }
\toprule
Parameter & Value & Role \\
\midrule
$\lambda_0,\lambda_1,\lambda_2$ & $1/3,1/3,1/3$ & Type identity, taxonomy, and angular embedding \\
$w_\tau,w_f$ & $0.5,0.5$ & Type and field terms within node substitution \\
$\beta,\gamma$ & $0.5,0.5$ & Positive base and configuration-sensitive null cost \\
$w_n,w_e$ & $0.5,0.5$ & Node and directed labelled edge objective terms \\
$\omega_k$ & $1$ for all $k$ & Fixed equal weighting over the 2,930-key field universe \\
Encoder & all-MiniLM-L6-v2 & Frozen type-description embedding table \\
Numerical tolerance & $10^{-12}$ & Bound pruning and incumbent update tolerance \\
Runner seed & 42 & Reproducible transformation and sampling seed \\
E1 repetitions & 5 per level & Twenty controlled transformations per base workflow \\
E5 audit sample & 50 graph IDs & Triangle evaluation sample \\
\bottomrule
\end{tabularx}
\end{table}

Table~\ref{tab:solver_coverage} summarizes the branch-and-bound-complete evaluation sets used by
the empirical analyses.  Completion is determined row by row through the stored
\texttt{used\_exact} flag; approximate upper bounds are not mixed into these
panels.

\begin{table}[t]
\centering
\caption{Branch-and-bound-complete evaluation sets used in the reported experiments.}
\label{tab:solver_coverage}
\begin{tabularx}{\textwidth}{@{} l r Y @{} }
\toprule
Artifact & B\&B-complete rows & Role in the analysis \\
\midrule
E1 graded transformation response & 32,093 & Complete level panel and monotonicity analysis \\
E2 accumulated edits & 8,031 & Complete one-to-five-edit panel \\
E3 weight ablation & 40,242 & Common-panel sensitivity analysis \\
E4 synthetic systems & 2,969 & One- versus three-perturbation discrimination \\
\bottomrule
\end{tabularx}
\end{table}

\section{Full empirical results}\label{app:experiments}

\subsection{Severity and accumulated edits}

Table~\ref{tab:e1_full} gives the row counts and duplicate-aware means behind
the main Experiment~1 figure.

\begin{table}[t]
\centering
\caption{Experiment~1 branch-and-bound-complete rows and duplicate-aware complete-panel
means.  Row means treat source-path copies separately; the 813-hash means are
the values plotted in Figure~\ref{fig:validation_core}.}
\label{tab:e1_full}
\begin{tabular}{@{} c r r r r @{} }
\toprule
Level & Stored rows & Complete hashes & Row mean & Equal-hash mean \\
\midrule
L0 surface & 8,420 & 813 & 0.000012 & 0.000012 \\
L1 fields & 8,420 & 813 & 0.000078 & 0.000078 \\
L2 attempted local & 8,063 & 813 & 0.347408 & 0.345896 \\
L3 major & 7,190 & 813 & 0.605812 & 0.584096 \\
\bottomrule
\end{tabular}
\end{table}

At the hash level, mean transitions are nondecreasing for 100.0\%, 99.754\%, and
84.010\% of the complete panel.  The first two transitions are effectively
universal.  The remaining overlap between L2 and L3 reflects the fact that the
levels contain heterogeneous local and multi-node transformation operators,
while the aggregate means preserve the intended severity ordering.

Experiment~2 is summarized in text because its complete-panel series has a single chain length dimension.  On the 796 canonical hashes with
all five branch-and-bound-complete steps, the equal-hash means for $k=1,\ldots,5$ are
0.175366, 0.327135, 0.471998, 0.572594, and 0.692323.  The pooled complete-panel
Spearman coefficient is 0.520337.  Within hashes, the mean and median
correlations are 0.842183 and 0.974679, and 96.36\% are positive.

\subsection{Weight ablation}

\begin{figure}[t]
\centering
\includegraphics[width=\textwidth]{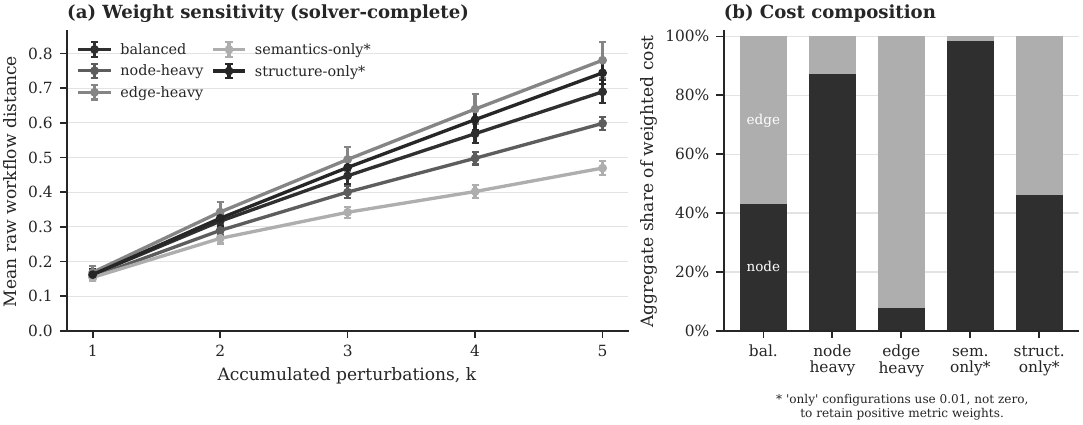}
\caption{Branch-and-bound-complete weight-ablation results on the common 788-hash panel.
Panel (a) equal-weights hashes and shows hash-bootstrap intervals.  Panel (b)
uses aggregate weighted-cost sums over the same hash-level cells, not the mean
of row-wise ratios.  The stored ``only'' labels retain 0.01 positive
weights and are diagnostic configurations, not literal removal of a graph
component.}
\Description{The left panel shows distance increasing with accumulated perturbations under five positive weight configurations. The right panel shows the aggregate node and edge cost shares for those configurations.}
\label{fig:ablation_weights}
\end{figure}

\begin{table}[t]
\centering
\caption{Weight specification, edit-count association, and aggregate cost
composition for Experiment~3 on the paired 788-hash panel.  Component shares
are aggregate weighted-cost shares computed from sums across those cells.}
\label{tab:e3_merged}
\begin{tabularx}{\textwidth}{@{} Y c c r r r @{} }
\toprule
Configuration & $(w_\tau,w_f)$ & $(w_n,w_e)$ & $\rho_S$ & Node & Edge \\
\midrule
Balanced & $(.50,.50)$ & $(.50,.50)$ & .5379 & 43.07\% & 56.93\% \\
Node-heavy & $(.50,.50)$ & $(.90,.10)$ & .5836 & 87.19\% & 12.81\% \\
Edge-heavy & $(.50,.50)$ & $(.10,.90)$ & .5074 & 7.75\% & 92.25\% \\
Field-dominant (stored \texttt{semantics\_only})
& $(.01,.99)$ & $(.99,.01)$ & .5091 & 98.48\% & 1.52\% \\
Type-and-edge (stored \texttt{structure\_only})
& $(.99,.01)$ & $(.50,.50)$ & .5398 & 46.21\% & 53.79\% \\
\bottomrule
\end{tabularx}
\end{table}

The stored names are imperfect: ``semantics only'' is primarily a
field-and-node weighting, while ``structure only'' retains a large type term.
Table~\ref{tab:e3_merged} therefore reports both the stored key and a literal
description.  Positive 0.01 weights preserve the formal positivity assumptions.

\subsection{Synthetic systems}

Table~\ref{tab:e4_full} reports the synthetic-system comparison after grouping
source-path copies.

\begin{table}[t]
\centering
\caption{Experiment~4 distributions after grouping source-path copies.  The
paired analysis uses the 757 common normalized-graph hashes.}
\label{tab:e4_full}
\begin{tabular}{@{} l r r r @{} }
\toprule
Synthetic system & B\&B-complete rows & Hashes & Equal-hash mean \\
\midrule
A: one perturbation & 1,581 & 804 & 0.307434 \\
B: three perturbations & 1,388 & 764 & 0.763241 \\
\bottomrule
\end{tabular}
\end{table}

On the paired hash subset, the corresponding means are 0.308961 and 0.761327,
medians are 0.204278 and 0.680785, and A has lower distance for 81.110\% of
hashes.  The one-sided paired Wilcoxon statistic is 38,475 with
$p=1.96244\times10^{-68}$.  The manuscript does not report the stored
\texttt{similarity} or \texttt{ees} columns: their generating formulas differ
from the normalizer in Equation~\eqref{eq:normalizer}.

\subsection{Triangle audit}

\begin{figure}[H]
\centering
\includegraphics[width=0.92\textwidth]{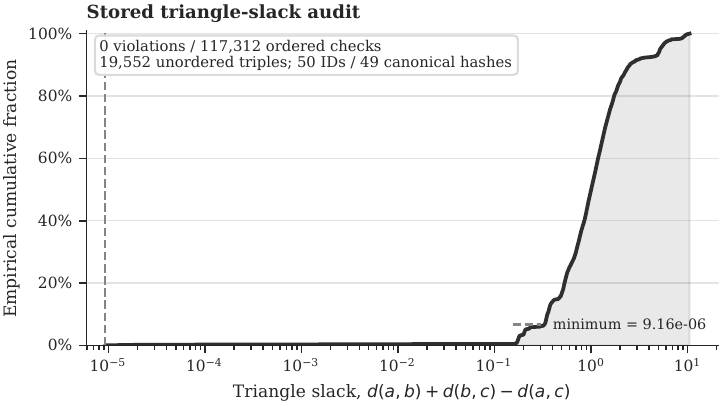}
\caption{Empirical CDF of the 117,312 stored ordered triangle slacks.  All
recorded slacks are positive.  The 50 path IDs represent 49 normalized-graph hashes
and only 55,366 distinct algebraic hash-level inequalities; ordered orientations
and duplicate paths are dependent.}
\Description{An empirical cumulative distribution of triangle slack remains entirely on the nonnegative side across 117,312 ordered checks.}
\label{fig:triangle_audit}
\end{figure}

Figure~\ref{fig:triangle_audit} and Table~\ref{tab:e5_full} summarize the
stored triangle-audit inventory.

\begin{table}[H]
\centering
\caption{Experiment~5 triangle-audit inventory for the stored run.}
\label{tab:e5_full}
\begin{tabular}{@{} l r @{} }
\toprule
Quantity & Recorded value \\
\midrule
Distinct graph IDs & 50 \\
Distinct normalized-graph hashes & 49 \\
Possible / completed graph pairs & 1,225 / 1,224 \\
Possible / completed unordered triples & 19,600 / 19,552 \\
Distinct-content pairs / triples represented & 1,176 / 18,424 \\
Stored ordered orientations & 117,312 \\
Distinct algebraic hash-level inequalities & 55,366 \\
Negative slacks at tolerance $10^{-9}$ & 0 \\
Minimum slack & $9.158491\times10^{-6}$ \\
Median slack & 1.00955 \\
99th percentile slack & 9.14001 \\
\bottomrule
\end{tabular}
\end{table}

The 50 path-identified graphs range from two to eight nodes.  The
formal distance is zero for identical canonical graphs.  The observed
same-content distance of $1.831698\times10^{-5}$ is an implementation-level
finite-precision residual of the same order as L0 and far below substantive
graph-change distances.  All
represented distinct-content pairs and triples are covered, and none of the
117,312 ordered triangle checks has negative slack at tolerance $10^{-9}$.

\section{Recovered provenance and reproducibility boundary}\label{app:provenance}

Table~\ref{tab:source_commits} records the recovered source snapshots and
their accepted corpus contributions.

\begin{table}[H]
\centering
\caption{Pinned public source snapshots and accepted corpus contribution.}
\label{tab:source_commits}
\begin{tabularx}{\textwidth}{@{} Y r r Y @{} }
\toprule
Source & Workflows & Nodes & Snapshot and licence evidence \\
\midrule
\path{enescingoz/awesome-n8n-templates} \cite{enescingozrepo}
& 291 & 5,645 & \texttt{e8fc6ee3bf00}; CC BY 4.0 \\
\path{newbie-learn-coding/n8n-Library} \cite{newbierepo}
& 2,348 & 36,606 & \texttt{79f0a501162f}; per-template/community licences \\
\path{Zie619/n8n-workflows} \cite{zie619repo}
& 2,066 & 30,806 & \texttt{ed8dd70c6959}; MIT \\
\bottomrule
\end{tabularx}
\end{table}

The three repositories are overlapping public collections rather than
independent samples.  Duplicate-aware normalized-graph hashes therefore define the unit
of analysis throughout the main results.

\texttt{ARTIFACT\_PROVENANCE.md} records SHA-256 values for the corpus,
registry, results, lineage manifest, summary tables, plotting scripts, and
figures used in the paper.  The source manifest maps all 4,705 evaluated IDs to
repository-relative paths and content hashes, enabling the duplicate-aware
analysis to be reproduced from the pinned snapshots.

\begingroup
\footnotesize
\sloppy
\bibliographystyle{unsrt}
\bibliography{references}
\endgroup

\end{document}